\documentclass[twocolumn]{article}
\usepackage[utf8]{inputenc}
\usepackage{textcomp,marvosym}
\usepackage[table,xcdraw]{xcolor}
\usepackage{sectsty}
\allsectionsfont{\sffamily}
\usepackage{abstract}

\usepackage{arydshln}
\usepackage{url}
\usepackage{rotating} 
\usepackage{listings}
\usepackage{hyperref}
\usepackage{float}
\usepackage{enumitem}
\setlist[enumerate]{noitemsep}
\usepackage{nameref,hyperref}
\usepackage{textgreek}
\usepackage{graphicx}
\usepackage{multirow}
\usepackage{listings} 

\renewcommand{\footnotesize}{\scriptsize} 
\usepackage[left=0.6in,right=0.6in,top=0.6in,bottom=0.8in]{geometry}
\usepackage{authblk}
\usepackage{booktabs} 
\usepackage[thinlines]{easytable} 
\usepackage{diagbox} 
\usepackage{slashbox} 
\usepackage{xcolor} 
\usepackage[backend=biber,style=nature,sorting=none]{biblatex}
\usepackage{tikz} 
\newcommand{\square}[1]{\tikz[baseline={([yshift=-0.8ex]current bounding box.center)}]{\filldraw[draw=black,fill=#1] (0,0) rectangle (0.25cm,0.25cm);}} 
\newcommand{\squarelarge}[1]{\tikz[baseline={([yshift=-0.8ex]current bounding box.center)}]{\filldraw[draw=black,fill=#1] rectangle (0.4cm,0.4cm);}} 

\providecommand{\keywords}[1]{\textbf{\textit{Keywords --}} #1}
\usepackage[font=small]{caption}

\begin{document}

\begin{refsection}

\title{\sffamily Phylogenetic reconstruction of the cultural evolution of electronic music via dynamic community detection (1975--1999) \rmfamily}

\date{}

\author[a,b,1]{\normalsize Mason Youngblood}
\author[c]{\normalsize Karim Baraghith}
\author[d]{\normalsize Patrick E. Savage}

\affil[a]{\scriptsize Department of Psychology, The Graduate Center, City University of New York, New York, NY, USA}
\affil[b]{\scriptsize Department of Biology, Queens College, City University of New York, Flushing, NY, USA}
\affil[c]{\scriptsize Department of Philosophy, DCLPS, Heinrich-Heine University, D{\"u}sseldorf, NRW, Germany}
\affil[d]{\scriptsize Faculty of Environment and Information Studies, Keio University SFC, Fujisawa, Japan}
\affil[1]{masonyoungblood@gmail.com}

\twocolumn[
\begin{@twocolumnfalse}
	
\maketitle
	
\vspace*{-20pt}
\begin{abstract}
\vspace*{-10pt}
		
Cultural phylogenies, or ``trees'' of culture, are typically built using methods from biology that use similarities and differences in artifacts to infer the historical relationships between the populations that produced them. While these methods have yielded important insights, particularly in linguistics, researchers continue to debate the extent to which cultural phylogenies are tree-like or reticulated due to high levels of horizontal transmission. In this study, we propose a novel method for phylogenetic reconstruction using dynamic community detection that explicitly accounts for transmission between lineages. We used data from 1,498,483 collaborative relationships between electronic music artists to construct a cultural phylogeny based on observed population structure. The results suggest that, although the phylogeny is fundamentally tree-like, horizontal transmission is common and populations never become fully isolated from one another. In addition, we found evidence that electronic music diversity has increased between 1975 and 1999. The method used in this study is available as a new R package called \textit{DynCommPhylo}. Future studies should apply this method to other cultural systems such as academic publishing and film, as well as biological systems where high resolution reproductive data is available, to assess how levels of reticulation in evolution vary across domains.

\\\\
		
\end{abstract}
\end{@twocolumnfalse}]

\section*{Introduction}

Historically, researchers have relied on phylogenetic comparative methods from biology to create cultural phylogenies, or ``trees'' of culture. These methods use differences and similarities in the cultural products of different populations (analogous to differences and similarities in DNA) to reconstruct the historical relationships between them. Traditional phylogenetic methods, which assume the tree-like structure typical of genetic evolution, have yielded critical insights, particularly in linguistics \cite{Pagel2007,Levinson2012,Bouckaert2012}. However, researchers have debated whether cultural phylogenies are fundamentally tree-like, or whether high levels of horizontal transmission lead to a more reticulated structure \cite{Boyd1997,BorgerhoffMulder2006,Gray2010,Rivero2016,Cabrera2017}. For example, biologist Stephen Jay Gould \cite{Gould1991} wrote that:

\begin{quote}
``Biological evolution is a bad analogue for cultural change [...] Biological evolution is a system of constant divergence without subsequent joining of branches. Lineages, once distinct, are separate forever. In human history, transmission across lineages is, perhaps, the major source of cultural change.''
\end{quote}

Certainly, this is an oversimplification. Horizontal transmission also frequently occurs in biology and sometimes provides problems for the practice of classification \cite{Doolittle2009}. Horizontal transmission is for instance the primary mechanism for the spread of antibiotic resistance in bacteria \cite{Gyles2014}, but it also occurs in vertebrates, invertebrates, and plants \cite{Crisp2015,Bock2010}. Nevertheless, Gould had a fair point. Although cultural traits evolve through a process of variation, selection and reproduction that results in observable fissions of cultural lineages, branches of the ``tree of culture'' can reunify later on and frequently do so, for instance in the case of shared practices or customs.

That being said, there is significant variation in reticulation across cultural domains \cite{Gray2010,Cabrera2017}, and depending on the conditions (e.g. co-inheritance of traits) horizontal transmission may or may not interfere with traditional phylogenetic reconstruction \cite{Nunn2006,Nunn2010,Currie2010}. In language evolution, for example, horizontal transmission is lower and phylogenetic relationships can be reliably reconstructed \cite{Greenhill2009}. In other domains, such as material culture, rates of horizontal transmission can be higher and more variable \cite{Tehrani2002,Jordan2003,Cochrane2010}, leading to phylogenies that clearly contradict the historical record \cite{Temkin2007}. For contemporary culture in the digital age \cite{Acerbi2020}, where rapid within-generational changes are the norm \cite{Youngblood2019a,Youngblood2019b}, the negative effects of horizontal transmission on phylogenetic signal are likely to be even more extreme. More recent advancements in network-based phylogenetics allow researchers to estimate reticulation \cite{Gray2010,Heggarty2010,Howe2011,Rivero2016,Mesoudi2017,Bouckaert2019}, but these methods are typically unrooted \cite{Tehrani2013a,Morrison2014a,Morrison2014b} (i.e. cannot be used to infer chronology \cite{Dunn2004}) and thus remain complementary to traditional phylogenetic reconstruction \cite{Tehrani2017}.

Additionally, outside of linguistics it can be extremely challenging to characterize complex cultural traits in a manner suitable for phylogenetic analysis \cite{Howe2011,Temkin2016}. In practice, this means that cultural phylogenies are often limited to very specific domains with variation that can be more easily characterized. For example, applications of phylogenetic methods in music have been restricted to traditional rhythmic patterns \cite{Toussaint2003,DiazBanez2004}, individual instruments \cite{Temkin2007}, the works of a single composer \cite{Liebman2012,Windram2014}, or folk music within a single region \cite{LeBomin2016}.

Given these limitations, it would be incredibly valuable to be able to construct large-scale phylogenies for complex cultural traits while explicitly accounting for and measuring horizontal transmission \cite{Nunn2006}. If phylogenies represent changes in population structure over time \cite{Tehrani2010,Velasco2013,Duda2016}, then one way forward might be to assess population structure from the bottom-up. This is, of course, a complicated proposition. Population structure in biology is typically determined by genetic variation driven by the combined effects of evolutionary processes such as recombination, mutation, genetic drift, demographic history, and natural selection. Unfortunately, population concepts are hardly discussed in the cultural evolutionary literature (for an exception see Scapoli et al. \cite{Scapoli2005}).

According to philosopher of biology Roberta Millstein's definition of a (biological) population, the ``boundaries of a population are those groupings where the rates of interactions are much higher within than without'' \cite{Millstein2010}. She calls this definition the ``causal interactionist population concept'', or CIPC for short. If agents are represented as nodes in a network graph, where links represent interactions relevant for reproduction (e.g. mating, transferring information), then populations are groups of agents that interact significantly more with one another than with other agents. In general, the rates of interactions between agents are lower between populations than within them, and this feature is precisely what gives them their (sometimes fuzzy) boundaries. It has recently been suggested to use this ``inner interactive connectivity'', i.e. cohesion in cultural population structure, as the population defining criterion in cultural evolution \cite{Baraghith2020}.

\begin{figure*}[ht]
\centering
\includegraphics[width=1\linewidth]{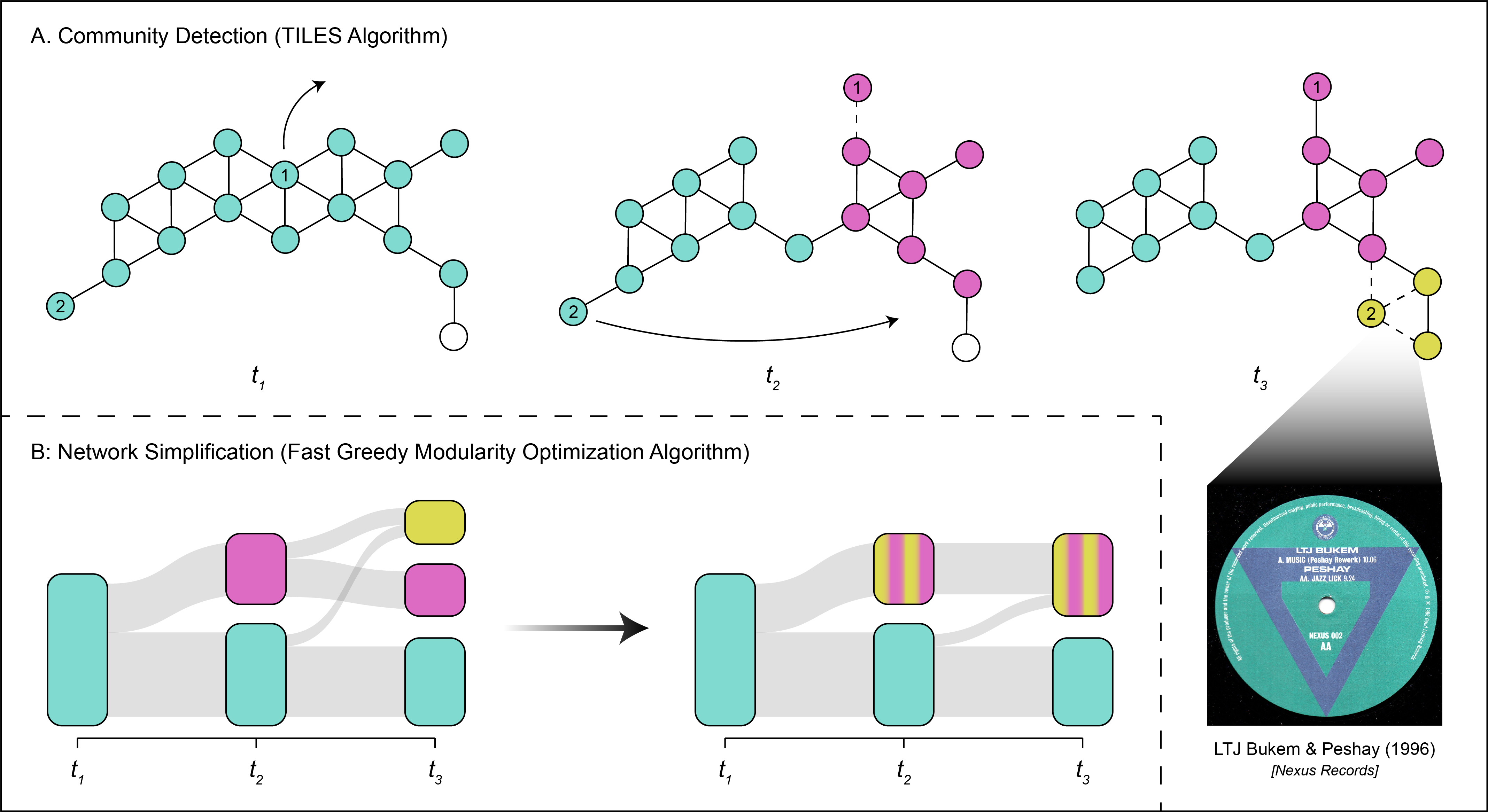}
\caption{A simplified visual summary of community detection using the TILES algorithm (A: top), and network simplification using the fast greedy modularity optimization algorithm (B: bottom). In A, arrows depict the movement of nodes (artists), dotted lines depict newly formed links (collaborations), and different colors represent different communities. Each link in the network corresponds to a collaborative release like the one shown in the bottom right (\url{https://bit.ly/3jvofD8}). Both core and peripheral artists are included, and artists in multiple communities are assigned to the largest one. Between $t_{1}$ and $t_{2}$, artist 1 breaks all but one of its previous ties. Since artist 1 is no longer in a triangle with those other artists, the network splits into two communities (green and pink). Between $t_{2}$ and $t_{3}$, artist 2 breaks its tie with the green community and establishes a collaboration with a peripheral member of the pink community and an unassigned artist. Since these three artists form a new triangle they are classified as a new community (yellow). In the left side of B, each node represents a community from A in the corresponding time point, and the links between communities represent the number of individuals moving between them. The right side of B is the same visualization after network simplification, where the pink and yellow communities have been compressed into a single population.}
\label{example}
\end{figure*}

According to this interpretation, populations represent a specific kind of ``nearly decomposable system'' \cite{Simon1962} in that they are emergent properties of the behavior of semi-independent agents. Naturally, agents in these kinds of nearly decomposable systems will be hierarchically organized on different levels, for example into communities, populations, and metapopulations. In the following, we propose a new method for phylogenetic reconstruction based on the CIPC that uses dynamic community detection to identify distinct populations and track how they change over time. We will treat different groups under investigation as populations or metapopulations, rather then as species. Deciding whether a member belongs to a specific community in this framework is not a question of whether she shares as many relevant hereditary traits as possible (like ``genes'' in biological evolution) with the other members of the grouping. Instead, what counts is the relative number of social interactions within (and outside) the community.

More specifically, we are using genres of electronic music as a test case. Although musicologists recognize that genres are generated by evolving communities of artists \cite{Lena2012,Klement2019}, previous attempts to quantitatively map genres have depended on listener habits \cite{Lambiotte2005}, instrument similarity \cite{Percino2014}, or sub-genre tags on streaming platforms \cite{Mauch2015}. By using artist co-release data (who collaborates with who) we can explicitly track how populations of artists, and the genres that they correspond to, evolve over time. This approach is similar to qualitative attempts at reconstructing music trees that rely on historical accounts of how artist communities grow, diverge, and influence each other over time \cite{Ishkur2000,Crauwels2016}. We chose to study electronic music because it is known for its rapid differentiation into competing genres and subgenres \cite{VanVenrooij2015}, particularly during the 1990s \cite{McLeod2001}. In addition, collaboration links between electronic music producers are already known to be important for cultural transmission \cite{Youngblood2019a,Youngblood2019b} and community structure \cite{Janosov2020}.

In brief, our method uses the TILES algorithm \cite{Rossetti2017} to identify communities in a dynamic network of artists, which are then clustered into populations using the fast greedy modularity optimization algorithm (Figure \ref{example}). TILES is an online algorithm in that it works with an ``interaction stream'' of nodes, in this case artists, and links, in this case collaborations. In other words, individuals enter and exit communities as they form and break relationships with other individuals. A node is considered to be a ``core'' community member if it forms a triangle with other community members, and a ``peripheral'' member if it is one link away from a core node. Community composition is recomputed throughout this process, each time that a new link enters or exits the network. Snapshots of the community composition are then collected at a regular time interval. We chose to use the TILES algorithm because it closely resembles the formalization of the CIPC proposed by Baraghith \cite{Baraghith2020}.

\definecolor{first}{RGB}{220, 218, 81}
\definecolor{second}{RGB}{222, 107, 195}
\definecolor{third}{RGB}{128, 123, 219}
\definecolor{fourth}{RGB}{221, 122, 89}
\definecolor{fifth}{RGB}{129, 218, 207}
\definecolor{sixth}{RGB}{136, 231, 91}
\definecolor{seventh}{RGB}{224, 224, 224}
\definecolor{eighth}{RGB}{122, 122, 122}

\begin{table*}
\centering
\scriptsize
\begin{tabular}{cccp{4.5cm}p{10cm}}
\toprule
& \textbf{N} & \textbf{CI} & \textbf{Top Styles} & \textbf{Top Artists} \\ \midrule
\square{first} 5 & 15,413 & 1.73 & Trance, House, UK Garage & Sash!, Vengaboys, 666, Paul van Dyk, Brooklyn Bounce, The Mackenzie, ATB, Fiocco, Da Hool, Nalin \& Kane \\ \midrule
\square{second} 3 & 10,891 & 1.45 & House, Euro House, Italodance & Cappella, Jon Of The Pleased Wimmin, Reel 2 Real, Sasha, 2 Unlimited, Corona, DJ Duke, DJ BoBo, Jeremy Healy, Dave Seaman \\ \midrule
\square{third} 8 & 8,228 & 4.61 & Synth-pop, Italo-Disco, Disco & Madonna, Depeche Mode, Pet Shop Boys, Black Box, Janet Jackson, Technotronic, Paula Abdul, Whitney Houston, New Order, Sandra \\ \midrule
\square{fourth} 1 & 8,152 & 1.05 & Techno, Hard House, Experimental & Mike Flores, DJ Spooky, AM/FM Alexander, George Centeno, Bill Laswell, Mad Professor, Poogie Bear, Mark Broom, Nemesis, Angel Alanis \\ \midrule
\square{fifth} 6 & 5,622 & 1.57 & Jungle, Drum n Bass, Hardcore & Grooverider, DJ SS, DJ Hype, Micky Finn, DJ Ratty, LTJ Bukem, Ellis Dee, Randall, DJ Phantasy, Kenny Ken \\ \midrule
\square{sixth} 4 & 5,243 & 1.98 & Hardcore, Happy Hardcore, Noise & Force \& Styles, Hixxy, Scott Brown, Bass Generator, Billy 'Daniel' Bunter, Buzz Fuzz, Eruption, Sy \& Unknown, DNA, DJ Fade \\ \midrule
\square{seventh} 2 & 3,885 & 4.87 & Disco, Synth-pop, New Wave & Commodores, Donna Summer, Giorgio Moroder, Patrick Cowley, Sylvester, Boney M., Kraftwerk, Gino Soccio, Yazoo, The Human League \\ \midrule
\square{eighth} 9 & 2,283 & 1.05 & Goa Trance, Psy-Trance, Progressive Trance & Astral Projection, Prana, Chakra, Quirk, MFG, S.U.N. Project, GMS, Pleiadians, X-Dream, ManMadeMan \\ \bottomrule
\end{tabular}
\caption{The size, cohesion index, top styles and top artists (calculated by weighted log-odds) from the eight largest populations in the phylogeny.}
\label{top_table}
\end{table*}

\section*{Methods}

\subsection{Data Collection}

All data used in the current study was collected from Discogs in April 2020. Discogs is a large, user-generated database of music releases that has better coverage of electronic music than other sources \cite{VanVenrooij2015,Bogdanov2017}. We chose to use the releases data rather than the masters data because it has better coverage of EPs, singles, remixes, and other smaller releases. First, we extracted all collaborative releases with the ``Electronic'' genre tag. Then, we retrieved the release IDs, release years, style tags, and artists from these releases, including any featured artists, remixers, and producers from the tracklists. Due to computational limitations, we only included releases from between 1970 and 1999. Collaborations were converted into an unweighted dynamic edgelist for input into TILES. More details about data collection and computational limitations can be found in the supporting information.

\subsection{Community Detection}

Community detection was conducted using the Python implementation of TILES \cite{Rossetti2017}. Each link persisted for 365 days and community composition was recorded at the beginning of each year. Links from collaborative releases were assigned random dates within the year they were released, so that changes in community membership were continuous.

The output of TILES is a set of community memberships for each year, where each individual may be part of multiple overlapping communities. In order to visualize the results, we required a single community membership for each individual. We assigned each individual in multiple communities to the community with highest network density, or the proportion of actual to possible connections. Ties were broken by group size. Both core and peripheral community members were included in the visualization and analysis.

\subsection{Network Simplification}

The dynamic community composition for all of the electronic releases, in which each node is a community of artists and each directed link represents artists moving between communities from year to year, was too large to visualize in its original form. To simplify the network we conducted a cluster analysis using the fast greedy modularity optimization algorithm \cite{Clauset2004} and merged nodes assigned to the same cluster within each year using the \textit{igraph} package in R \cite{Csardi2006}. Links were combined to represent the total number of artists moving between each set of merged nodes. The simplified network can be thought of as a hierarchical network with two layers. Each node represents a population of communities (identified by TILES) which are further separable into individual artists.

\subsection{Horizontal Transmission}

To measure horizontal transmission between two populations, we divided the total number of weighted links between populations by the sum of the number of internal weighted links from each population. In other words, we calculated the percentage of links in the sub-network that flow between the two populations. Since collaborative links between artists are meaningful for cultural transmission \cite{Youngblood2019a}, the level of collaboration between two populations likely reflects the level of cultural transmission between them. This simple metric is consistent with the common definition of horizontal transmission as the level of cultural transmission between extant populations \cite{BorgerhoffMulder2006,Greenhill2009,Currie2010}, and reflects the conceptualization of horizontal transmission in biological phylogenetics. Similarly, we define vertical transmission as cultural transmission occurring within populations. The original definition of horizontal transmission proposed by Cavalli-Sforza and Feldman, which was simply cultural transmission occurring within the same generation \cite{Cavalli-Sforza1981}, is insufficient here because it does not take population structure into account.

In addition, we calculated the cohesion index (CI), or the ratio of internal to external links, for each individual population \cite{Baraghith2020}. Although this application of the CI differs from its original formulation, as populations in the dynamic network have a temporal dimension, it is still an intuitive metric for the degree to which a population is distinct from other populations in the phylogeny. Since the weighted links are summed during both community detection and network simplification, the CI can be calculated at any level of the hierarchical network.

\subsection{Visualization Algorithm}

The phylogeny was visualized with a sankey plot using the \textit{sankeyD3} package in R \cite{Breitwieser2017}. This method expands on recent work by Mall et al. \cite{Mall2015} and others \cite{Rosvall2010,Wu2016}. As you go from left to right in the visualization, each node is a population of communities identified by TILES in the corresponding year on the \textit{x}-axis. The links depict movement between populations over time, where thickness corresponds to population size. Only the largest connected component of the network was plotted to maximize the clarity of the visualization. The layout of the nodes in the sankey plot is algorithmically generated by (1) minimizing the overlap between links and (2) maximizing the horizontal alignment of nodes. In other words, the visualization algorithm maximizes both the clustering of nodes and the continuity of links. The distances between and the positions of populations do not convey information like they do in a standard phylogeny, so the specific \textit{y}-axis location of each node is not meaningful.

\subsection{Topic Modelling}

To determine whether or not populations of artists correspond to distinct subgenres, we compiled the style tags for all releases from each population and conducted Latent Dirichlet Allocation (LDA) topic modelling. LDA models are Bayesian mixture models that take large corpora of words from different sources and identify clusters of words, or ``topics'' \cite{Grun2011}. Each source is assigned a distribution of topics, one of which is identified as the most distinctive to that source. In this case, we looked for clusters of style tags that were distinctive to each population. Style tags from Discogs have been successfully used to quantitatively identify genres in the past \cite{Mauch2015}, which is unsurprising given that their purpose is subjectively denoting genres. Releases by artists from multiple populations (42.5\%) were excluded from the topic modelling so that the samples were independent from one another. We used all style tags for each group rather than unique style tags to account for frequency of use. The LDA model was fitted using variational expectation maximization using the R package \textit{topicmodels} \cite{Grun2011}. \textit{k}, or the target number of topics, was set to the number of populations.

\section*{Results}

We analyzed 1,498,483 collaborative relationships between 93,831 artists from 53,581 different electronic music releases between 1970 and 1999. TILES identified 8,354 communities, of which we visualized the largest connected component. This component contained 90.2\% of the communities, appearing between 1975 and 1999. Network simplification reduced the dynamic community composition to 72 populations with a modularity score of 0.65. The geographic distribution of releases from these populations can be seen in Figure \ref{world_map}.

The resulting phylogeny can be seen in Figure \ref{phylo}. Each node is a set of communities identified by TILES, which are further separable into individual artists. Nodes with the same color belong to the same population. Each link corresponds to the number of artists moving between nodes. In the interactive version of Figure \ref{phylo} (\url{https://masonyoungblood.github.io/electronic_music_phylogeny.html}) you can zoom, navigate, and hover over nodes for the three most distinctive style tags and 10 most distinctive artists from each population (identified by weighted log-odds, calculated using uninformative Dirichlet priors), and hover over links for the number of individuals moving between nodes. Details about the eight largest populations in the phylogeny can be seen in Table \ref{top_table}.

\begin{table}
\centering
\footnotesize
\begin{TAB}(e,0.6cm){|c|c:c:c:c:c:c:c:c|}{|c|c:c:c:c:c:c:c:c|}
& \squarelarge{first} 5 & \squarelarge{second} 3 & \squarelarge{third} 8 & \squarelarge{fourth} 1 & \squarelarge{fifth} 6 & \squarelarge{sixth} 4 & \squarelarge{seventh} 2 & \squarelarge{eighth} 9\\
\squarelarge{first} 5 & \diagbox[width=0.6cm,height=0.6cm]{}{} & 12.3\% & 0.0\% & 12.9\% & 2.4\% & 4.0\% & 0.0\% & 3.9\% \\
\squarelarge{second} 3 & & \diagbox[width=0.6cm,height=0.6cm]{}{} & 4.3\% & 7.0\% & 7.4\% & 2.7\% & 0.0\% & 3.0\% \\
\squarelarge{third} 8 & & & \diagbox[width=0.6cm,height=0.6cm]{}{} & 0.0\% & 2.7\% & 0.1\% & 5.3\% & 0.0\% \\
\squarelarge{fourth} 1 & & & & \diagbox[width=0.6cm,height=0.6cm]{}{} & 5.2\% & 5.1\% & 0.0\% & 3.6\% \\
\squarelarge{fifth} 6 & & & & & \diagbox[width=0.6cm,height=0.6cm]{}{} & 6.4\% & 0.0\% & 1.8\% \\
\squarelarge{sixth} 4 & & & & & & \diagbox[width=0.6cm,height=0.6cm]{}{} & 0.0\% & 2.3\% \\
\squarelarge{seventh} 2 & & & & & & & \diagbox[width=0.6cm,height=0.6cm]{}{} & 0.0\% \\
\squarelarge{eighth} 9 & & & & & & & & \diagbox[width=0.6cm,height=0.6cm]{}{} \\
\end{TAB}
\caption{The level of horizontal transmission between the eight largest populations in the phylogeny. Each value represents the percentage of links in the sub-network that flow between the two populations. In other words, we divided the total number of weighted links between populations by the sum of the number of internal weighted links from each population.}
\label{ht_table}
\end{table}

Population \#5 (yellow) includes a diverse array of four-on-the-floor dance music but leans towards trance. Most of the top artists are from Germany. Population \#3 (pink) also includes a diverse array of dance music but leans towards house. The top artists are from America, England, and all over Europe. ``Euro House'', the second top style in this population, is associated with the radio-friendly popular dance music that rose to prominence in Europe in the mid 1990s. Population \#8 (purple) includes artists like Madonna, Pet Shop Boys, and Technotronic that were incorporating electro, hip-hop, and techno into pop and rock music in the late 1980s and early 1990s. Population \#1 (orange) includes a diverse array of artists from American and Europe making techno, hard house, and experimental music. Population \#6 (blue) is the branch of the ``hardcore continuum'' of high-BPM rave genres that includes jungle and drum n bass. Most of the top artists are from London and the surrounding suburbs. Population \#4 (green), on the other hand, is the heavier branch of the ``hardcore continuum'' that includes happy hardcore, gabber, and noise. Most of the top artists are from rural England and Scotland. Population \#2 (light grey) includes both disco and synth-pop throughout the 1980s, including Giorgio Moroder and Donna Summers who transformed dance music with their 1977 hit ``I Feel Love''. Population \#9 (dark grey) corresponds to Goa trance and psy-trance. Goa trance first emerged in India in the early 1990s and became known as psy-trance once it reached Europe and the rest of the world. Most of the top artists are from Europe and the Middle East. The most distinctive artists from each population do not necessarily represent the most innovative or influential artists from the subgenres that they correspond to. For example, the originators of Detroit techno music (e.g. Juan Atkins, Derrick May, and Kevin Saunderson \cite{Sicko2010}) do not appear in the most distinctive artists from Population \#1 (orange), and the originators of Chicago and New York house music (e.g. Frankie Knuckles and Larry Levan \cite{Reynolds2012}) do not appear in the most distinctive artists from Population \#3 (pink). The rise and fall of each of these lineages, as well as the interactions between them, are intuitive and broadly consistent with the historical record \cite{Sicko2010,Reynolds2012,Collins2013}.

\begin{figure}
\centering
\includegraphics[width=0.9\linewidth]{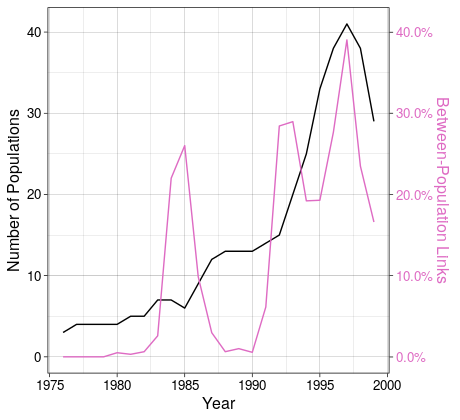}
\caption{The number of populations identified in each year (left \textit{y}-axis), as well as the percentage of between-population links between each year and the previous year (right \textit{y}-axis). The percentage of between-population links should be treated as a global measure of horizontal transmission.}
\label{time}
\end{figure}

The level of horizontal transmission between the eight largest populations in the phylogeny can be seen in Table \ref{ht_table}. 20.9\% of links in the phylogeny are between rather than within populations, indicating that while horizontal transmission is common the phylogeny still has a fundamentally branching structure. Interestingly, both the number of populations and the percentage of between-population links has increased over time (Figure \ref{time}). The large spike in the percentage of between-population links between 1984 and 1986 is likely due to the large movement of artists from Population \#2 (light grey) to Population \#8 (purple), which appears to be caused by an influx of new artists collaborating with existing artists and disrupting the original population structure.

The LDA topic model identified 48 unique topics across 69 of the populations in the simplified network. Three populations (\#53, \#57, \#62) were excluded from the topic model because they did not have any unique releases. The $\alpha$ value of the fitted model, which gets smaller as more topics correspond one-to-one with groups, was 0.019. 64.6\% of topics corresponded to a single population, 27.1\% corresponded to two populations, and 8.3\% corresponded to three populations. This indicates that populations of artists have distinctive style tags associated with their releases, and likely represent distinct subgenres of electronic music. The top 10 terms from each topic assigned to the eight largest populations in the phylogeny can be seen in Figure \ref{lda_viz}.

\begin{figure}
\centering
\includegraphics[width=0.9\linewidth]{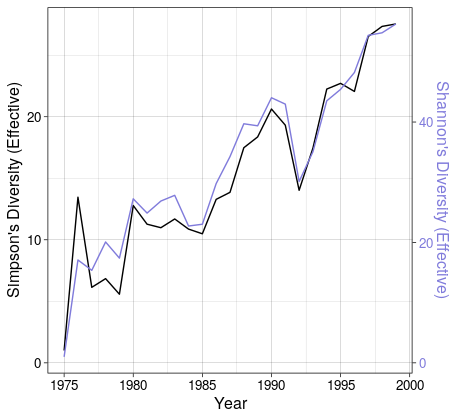}
\caption{The effective number of styles from each year calculated using Simpson's (left \textit{y}-axis) and Shannon's (right \textit{y}-axis) diversity indices.}
\label{diversity}
\end{figure}

The style tag diversity from each year, calculated with all releases in the phylogeny using the Simpson and Shannon diversity indices, can be seen in Figure \ref{diversity}. We used the effective number of styles, or the number of equally-abundant styles required to get the same diversity index \cite{Jost2006}, because it scales linearly and has been used in previous work on musical diversity \cite{Mauch2015}. Between 1975 and 1999 the diversity of styles tags in the phylogeny has significantly increased.

\section*{Discussion}

By applying dynamic community detection methods to an exhaustive dataset of electronic music releases spanning three decades, we have constructed a cultural phylogeny built explicitly from population structure. Most importantly, we found that the phylogeny has a tree-like structure, in that the majority of links in the phylogeny are within lineages (79.1\%) and the number of populations increases over time. However, reticulation is ubiquitous and populations never become fully isolated from another. Additionally, the level of horizontal transmission between populations increased between 1975 and 1999. This is consistent with anecdotal accounts \cite{Lindop2011} and another recent quantitative study that found that hybridization between genres increased over the same period, using genre tags from music databases \cite{Gagen2019}. Taken together these results indicate that although communication technologies like the internet have increased rates of horizontal transmission between human populations \cite{Carrignon2019}, allowing music producers to collaborate and influence one another despite significant geographical distance \cite{Youngblood2019a}, the cultural evolution of music is still a fundamentally branching process.

We also observed a significant increase in style tag diversity. In combination with population diversification, this indicates that electronic music diversity has increased between 1975 and 1999. This result contradicts two recent studies on the cultural evolution of popular music, which concluded that musical diversity has either remained relatively constant \cite{Mauch2015} or declined \cite{Serra2012} in recent decades. Although it is possible that diversification patterns in popular music are different than in contemporary music as a whole, both studies have some important limitations that reduce their generalizability. Mauch et al. \cite{Mauch2015} only used data from the Billboard Top 100. Estimates of diversity based on the Billboard charts alone will inevitably underestimate the diversity in music being produced at all levels of an ever-expanding industry. Although Serr{\`{a}} et al. \cite{Serra2012} used a more extensive dataset their conclusions are based solely on pitch transitions, timbre, and loudness, measures that only capture a fraction of musical variation and are less relevant for recent musical innovations (e.g. experimental, noise, drum-based, and lyric-based music). That being said, the quantitative audio analysis methods used by both Mauch et al. and Serr{\`{a}} et al. \cite{Serra2012,Mauch2015} better capture patterns of musical diversity than style tags. Future studies should combine quantitative audio analysis with our process-based phylogenetic approach to better understand the diversification of contemporary music.

The tree-like structure of the phylogeny is likely due to the presence of transmission isolating mechanisms, or TRIMS \cite{Durham1992}, that reduce levels of transmission between cultural lineages. TRIMS are comparable to intrinsic reproduction barriers that exist between members of different biological species. ``Intrinsic'' because they should be separated conceptually from extrinsic reproductive barriers that separate biological populations (e.g. two snakes of the same species being separated by a mountain chain or a population of fish being isolated in a lake after a drought). For example, an empirical study by Tehrani et al. \cite{Tehrani2013b} revealed that knowledge of textile production in Iranian tribal groups, who still live in a very traditional way, is primarily vertically transmitted because of social norms around endogamous marriage and gender. Mothers teach craft skills to their daughters, who almost always marry within their tribal group, and sharing of designs is restricted to other women in the same community.

In contemporary music there also appear to be TRIMS that reduce cultural transmission between groups. Individuals in ``underground'' electronic music communities often define themselves in opposition to mainstream music culture, and are deeply invested in the shared identity of their community \cite{Thornton1995,Lindop2011}. The common underlying concern, that too much mainstream recognition could reduce the longevity and undermine the integrity of music communities, is actually supported by previous research on electronic music in the UK \cite{VanVenrooij2015}. Artists who receive mainstream recognition are sometimes subject to criticism \cite{Thornton1995,Noys1995,Hesmondhalgh1998}, and may even adopt aliases to reinforce their commitment to the ``scene'' \cite{Formilan2020}. Artists also appear to intentionally adopt styles that sound distinctive relative to more popular artists \cite{Klimek2019}, which could be further reinforced by conformity bias within groups \cite{Youngblood2019b}. Communal experiences in clubs and raves, sometimes supplemented with drug use, also enhance social bonding \cite{Hutson2000,StJohn2006,Savage2020} and reinforce community boundaries \cite{Kavanaugh2008}. Some scholars have even posited that obscure subgenre names and other forms of jargon function in ``maintaining clear boundaries that define in-group/out-group relations'' \cite{McLeod2001}. We hypothesize that these social norms act as TRIMS by enhancing the longevity and cohesion of music communities, and thus reducing the likelihood that they are integrated into larger and more popular genres \cite{Lena2012}.

Importantly, many populations evolve independently and only later connect to the larger phylogeny. In some cases this is likely due to the fact that external connections to other genres are not shown, but in other cases it could represent independent evolution via shared technology. For example, the emergence of electronic music in India in the 1970s, most notably in Bollywood (\#15 in Figure \ref{phylo}), was driven by the introduction of novel synthesizer technologies rather than the influence of American and European artists \cite{Pandey2019,Purgas2020}. Additionally, the level of reticulation that we observed for electronic music has likely been enhanced by communication technologies and a globalized music industry, and is not necessarily reflective of traditional music. Diversity in Gabonese folk music, for example, appears to show signatures of vertical transmission \cite{LeBomin2016}, and phylogenetic studies indicate that folk music variation can be relatively conserved within genetic lineages \cite{Pamjav2012,Brown2013}. Unfortunately, our method can only be used in cases where detailed collaboration data is available, which is unlikely to be the case for any form of traditional music.

Several limitations of this study need to be highlighted. Firstly, the phylogeny was exclusively constructed from collaboration links, which do not account for all cultural transmission. For example, contemporary artists routinely draw inspiration from recorded music, live events, new technologies, etc. Thus, the phylogeny really only captures the ``core'' of these cultural lineages \cite{Boyd1997}. Secondly, the data was limited to releases tagged as ``electronic'' on Discogs, so external connections to other genres (e.g. rock and pop music) are not shown.

Our framework also has the potential to improve music recommendations on streaming platforms like Spotify. Currently, Spotify's recommendation algorithms are primarily based on listener habits \cite{Johnston2019}. In other words, related artists are identified based on what their fans are also listening to. Spotify's algorithms have been observed to reduce the diversity of listeners' music consumption \cite{Anderson2020}, and have the potential to further compound existing inequalities in the music industry based on popularity \cite{Abdollahpouri2020a,Abdollahpouri2020b}, gender \cite{Werner2020}, and other factors \cite{ODair2020}. Supplementing recommendation algorithms with data on collaborative population structure would allow for the identification of artists from the same community independently of listeners' biases.

Future research should apply this method to high resolution data from other cultural domains (e.g. co-citation patterns in academic fields \cite{Youngblood2018}, institutional membership in the arts \cite{Fraiberger2018}, or the composition of Hollywood film crews \cite{Tinits2020}) to determine whether the observed level of reticulation in this study is typical of other contemporary cultural systems. Additionally, this method is theoretically generalizable to biological evolution as well. For example, researchers that study the evolution of insects in the lab could use motion tracking \cite{Crall2015} to estimate population structure from mating events, and then measure how genetic changes map onto lineages over time. In some long-term study systems, such Darwin’s finches, researchers have collected enough detailed mating data to reconstruct population structure in the wild. Applying this method where interbreeding between closely related species has led to hybrid speciation and clear genetic changes \cite{Lamichhaney2018} could provide new insight into the process of reproductive isolation.

\section*{\large Data \& Code Availability Statement}

\begin{sidewaysfigure*}
\centering
\includegraphics[width=\linewidth]{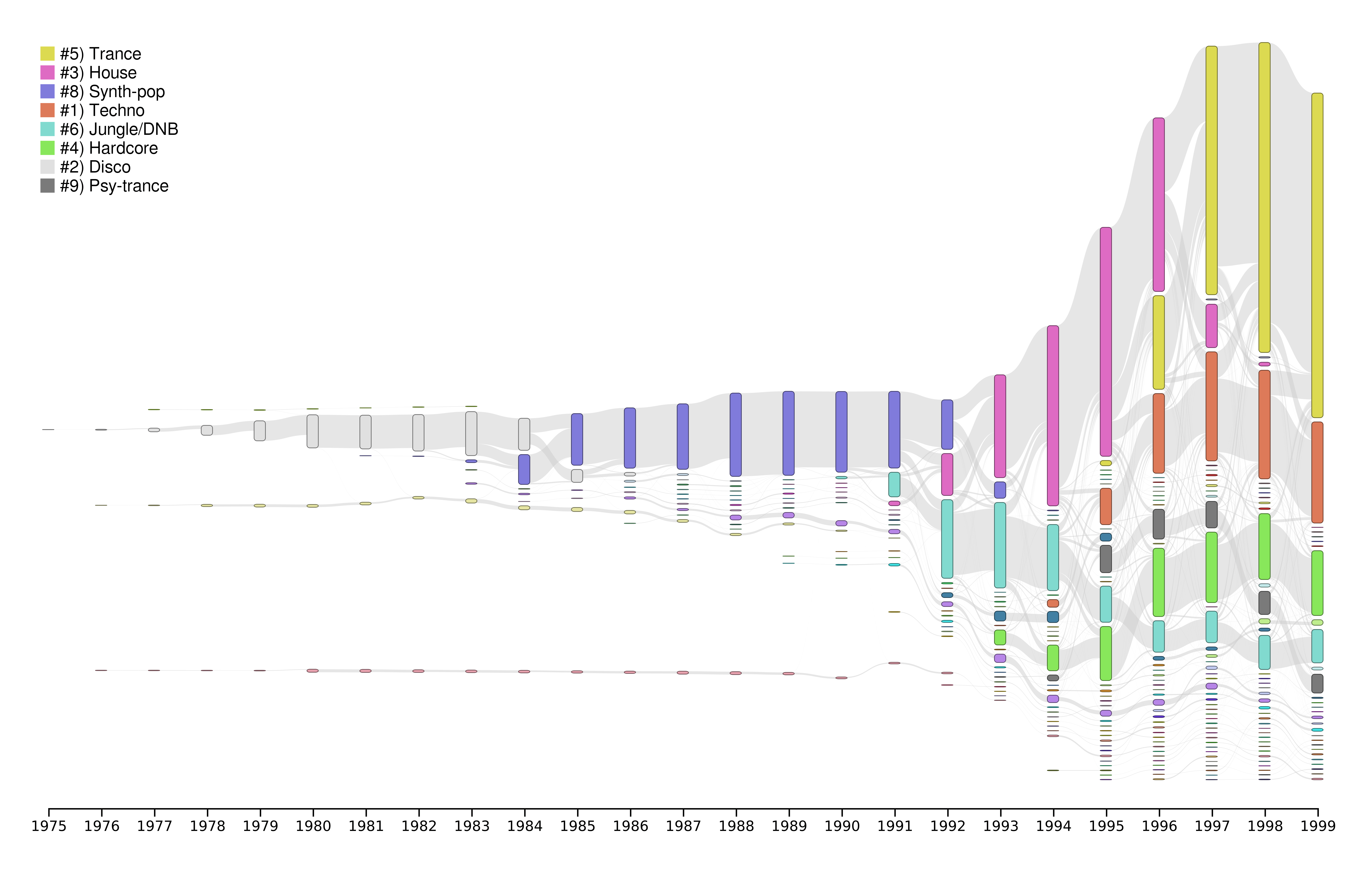}
\caption{A phylogeny based on the dynamic community composition of electronic music over time. Each node is a set of communities, and each link corresponds to the number of artists moving between nodes. Nodes with the same color belong to the same population, and the height of each node corresponds to the number of artists. The visualization algorithm maximizes both the clustering of nodes and the continuity of links. The population names in the legend are simplified versions of the more detailed classifications in the results. Interactive version: \url{https://masonyoungblood.github.io/electronic_music_phylogeny.html}}
\label{phylo}
\end{sidewaysfigure*}

The method used in this study is available as a new R package called \textit{DynCommPhylo} on Github (\url{https://github.com/masonyoungblood/DynCommPhylo}). The ``example'' subfolder of the R package includes our processed data and analysis script. The interactive version of Figure \ref{phylo} is also available on Github (\url{https://masonyoungblood.github.io/electronic_music_phylogeny.html}). The raw data used in this study came from the April 2020 XML archive of releases on Discogs (\url{http://bit.ly/DiscogsApril2020}).

\section*{\large Acknowledgments}

This research was supported, in part, under National Science Foundation Grants CNS-0958379, CNS-0855217, ACI-1126113 and the City University of New York High Performance Computing Center at the College of Staten Island.

\renewcommand*{\bibfont}{\scriptsize}
\printbibliography[title=\large References]

\end{refsection}

\newpage
\onecolumn

\begin{refsection}

\begin{center}
	\sffamily\LARGE Supporting information\rmfamily
\end{center}

\setcounter{figure}{0}
\setcounter{table}{0}
\setcounter{subsection}{0}
\renewcommand{\thetable}{S\arabic{table}}
\renewcommand{\thefigure}{S\arabic{figure}}

\subsection{Data Collection}

XML parsing and data management was conducted in BaseX (v9.3.2) and R (v3.6.1), using the follow XQuery commands:

{\small
\begin{lstlisting}[breaklines=true]
//release[genres/genre/text() = "Electronic"]/string-join((concat('"', @id, '"'), concat('"', master_id, '"'), concat('"', released, '"'), concat('"', replace(string-join((formats/format/descriptions/description), ', '), '"', ''), '"'), concat('"', string-join((styles/style), ', '), '"'), concat('"', string-join((artists/artist/id, tracklist/*/artists/artist/id), ', '), '"'), concat('"', country, '"')), '&#09;')
\end{lstlisting}}

The ``extra artists'' tag was not included, as it is often used for non-collaborators such as label executives, managers, and audio engineers. Releases tagged as compilations were excluded from the analysis, and the Discogs placeholder tags were removed (194: various artists, 355: unknown artist, 118760: no artist).

When multiple versions of a release were available, we used the one with the most credited artists and assigned it the earliest release year. Since Discogs also includes re-releases and promos, we manually checked the release years from a random sample of 500 releases to assess data quality. 98.4\% of releases had the correct year. Of the 1.6\% that were incorrect the majority were only one or two years off (because of incomplete artist data, bootleg versions, etc.). Only one was off by a significant amount of time (12 years), because the re-release was the only version labelled as ``electronic''. Rare releases with very inaccurate release years are unlikely to influence community detection because the collaborators of the artists are no longer part of the network.

\begin{figure}[H]
\centering
\includegraphics[width=1\linewidth]{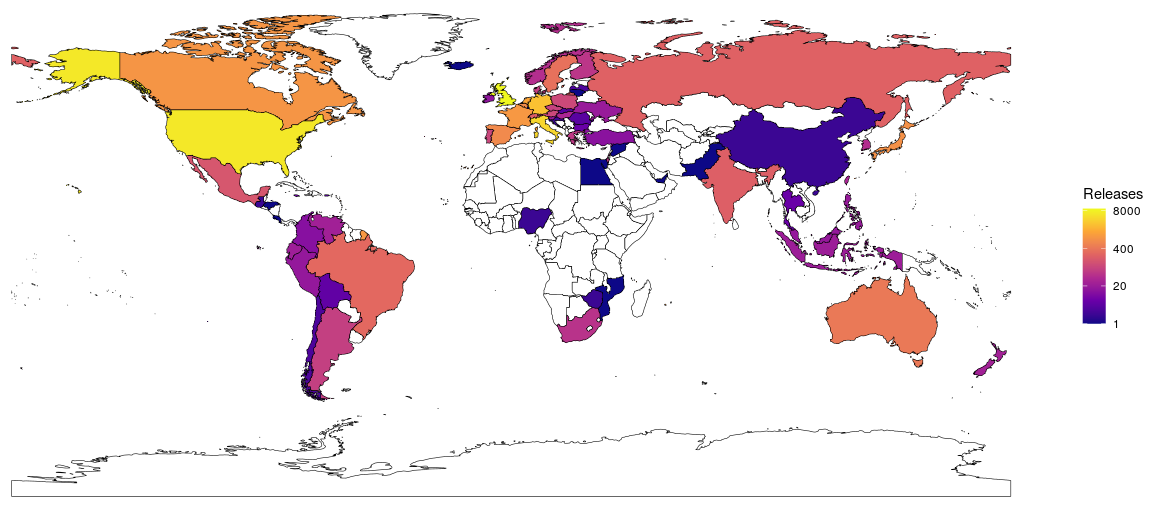}
\caption{The log-transformed geographic distribution of releases in the phylogeny, excluding the 3.6\% of releases with either missing data or a region instead of a country. A total of 72 countries are represented, with 50.3\% of releases coming from the US and UK.}
\label{world_map}
\end{figure}

\subsection{Computational Limitations}

The TILES algorithm was run on the Karle symmetric multiprocessor (72 cores, 768 GB memory) at the City University of New York High Performance Computing Center at the College of Staten Island. Analyzing all data from 1970-1999 took 79 hours, with 1998-1999 accounting for about 30 hours of that time. Analyzing a random subsample of 50\% of the data took only 1 hour, with 1998-1999 accounting for about 15 minutes of that time. This means that the runtime for the final year increased 120x with only 2x data. As the yearly data on Discogs throughout the 2000s and 2010s is roughly double or triple what it was in the 1990s, including more time with the full dataset was not really feasible and with the 50\% subsample would have only given us a few more years of results. As such, we chose to restrict the analysis to 1970-1999.

That being said, the results with the 50\% subset were very similar to the results we got with the full dataset. The eight largest populations were still present and had roughly the same level of horizontal transmission between them. The only major differences were in the smaller populations. For example, the independent evolution of electronic music in India (\#15 in Figure \ref{phylo}) completely disappeared. Future studies might explore using a random subsample of data to increase the computational efficiency of TILES.

\subsection{LDA Results}

\begin{figure}[H]
\centering
\includegraphics[width=0.75\linewidth]{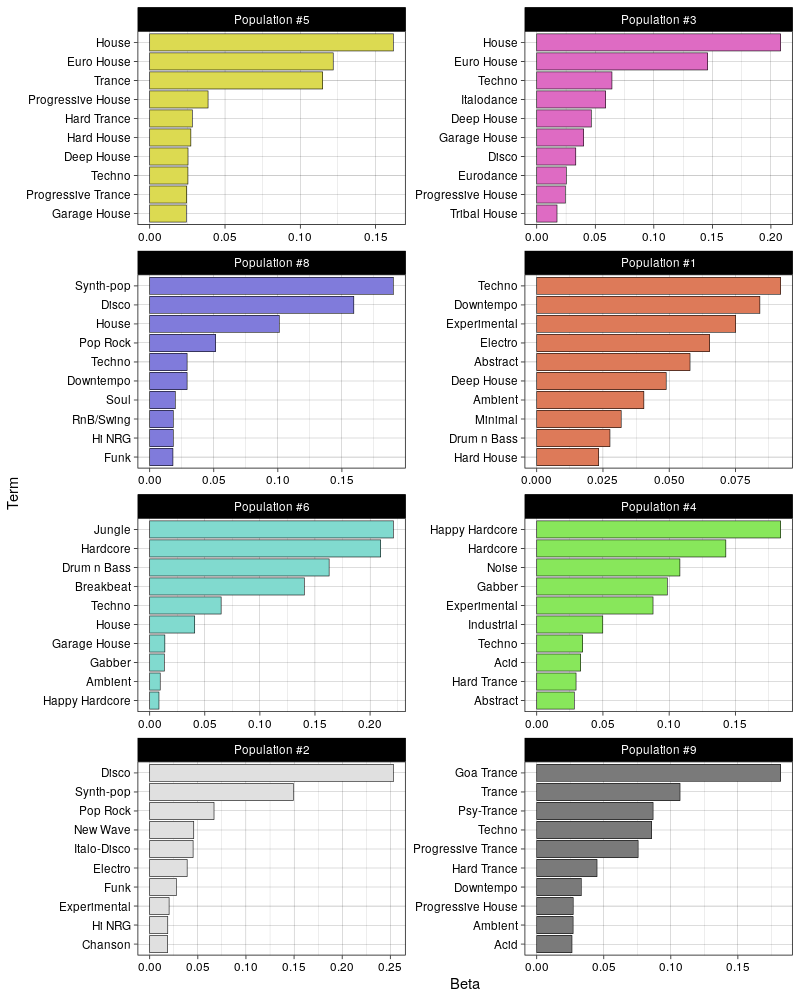}
\caption{The top 10 terms (ranked by $\beta$) from each topic assigned to the eight largest populations in the phylogeny. $\beta$ is the probability of a style coming from each topic.}
\label{lda_viz}
\end{figure}

\end{refsection}

\end{document}